# Polytropic bulk viscous cosmological model with variable G and Λ

[1]**Rishi Kumar Tiwari**, [2]**Mukesh Sharma** and [3]**Sonia Sharma**

[1]Department of Mathematics, Govt. Model Science College, Rewa (M.P.) India

[2], Department of Applied Sciences, R. I. E. I. T., Railmajra, SBS Nagar (Pb.) India

[3] Department of Mathematics, Rayat Polytechnic College, SBS Nagar (Pb.) India

Email: soniamathematics@yahoo.co.in

**Abstract**

We consider a Bianchi type-I Polytropic bulk viscous fluid cosmological model with variable G and Λ. To get a deterministic model, it is assumed that $\frac{\dot{A}}{A} = \frac{\dot{B}}{B} = \frac{k_1}{t^n}$ and $\frac{\dot{C}}{C} = \frac{k_2}{t^n}, \eta = \eta_o \rho^\alpha, \overline{p} = k\rho^\gamma$, where $\overline{p}$ is the pressure, ρ is the energy density, η is the coefficient of bulk viscosity, α, k, γ and $\eta_o$ are constants, H is Hubble constant, $\frac{k_1}{2} = k_2$ where $k_1 > 0$, $k_2 > 0$. A variety of solutions are presented. The physical significance of the cosmological models have also been discussed.

1. Introduction

Cosmology is the study of largest-scale structures and dynamics of our universe and is deals with subjects regarding its origin and evolution. Cosmology involves itself with studying the motions of the celestial bodies. The twentieth century advances made it possible to speculate about the origins of the universe and allowed scientists to establish the Big Bang as the leading cosmological model, which most cosmologists now accept as the basis for their theories and observations. Bianchi type-I space time is the simplest generalizations of Friedmann-Robertson-Walker (FRW) flat models. There is significant observational evidence that the expansion of the universe is undergoing late time acceleration (Perlmutter et al.1997,1998, 1999; Riess et al. 1998; Allen et al. 2004; Peebles et al. 2003; Padmanabhan 2003; Lima 2004)

The cosmological constant problem is very interesting. Recent observations indicate that $\Lambda \sim 10^{-55}$ cm$^{-2}$ while the particle physics prediction for Λ is greater than this value by a factor of order $10^{120}$. In modern cosmological theories, the cosmological constant remains a focal point of interest. The bulk viscosity associated with grand unified theory phase transition can lead to the inflationary universe scenario. A wide range of observations now suggests compellingly that the universe possesses a non-zero cosmological constant (Riess et al. 2004). H. Zhang et al.(2005) studied the Friedmann cosmology on codimension 2 brane with time dependent tension. In this model the effective cosmological constant is free of the absolute value of the brane tension. Tiwari and Sonia (2011) investigated the non-existence of shear in Bianchi type-III string bulk viscous cosmological model in general relativity. Wang (2003, 2004, 2005, 2006) discussed the solutions of Bianchi type I-IX cosmological models for string clouds. Also several aspects of viscous fluid cosmological models in early universe have been extensively investigated by many authors. Tiwari and Sonia (2011) investigated the Bianchi type-I string cosmological model with bulk viscosity and time dependent Λ term. Zeyauddin and Saha (2013) investigated the Bianchi type-V bulk viscous cosmological models with particle creation in general relativity. Beesham et al. (2000) investigated the Bianchi type-I anisotropic bulk viscous fluid cosmological model with variable G and Λ, where bulk viscosity $\eta=\eta_o\rho^n$, $\eta_o \geq 0$, ρ is the energy density and deriving Friedmann type equation $3H^2 = G\rho + \sigma^2 + \Lambda$, with σ being the shear and H the Hubble constant. Tiwari et al. (2009, 2010) and Tiwari et al. (2008, 2009, 2010) studied the isotropic and anisotropic cosmological models by taking $\Lambda \propto H, \Lambda \propto R^{-m}$ with H being the Hubble parameter, R the scale factor and m is a constant. Also Tiwari et al. (2011) investigated the LRS Bianchi type-II cosmological model with a decaying lambda term and Tiwari (2010) investigated the Bianchi type-I cosmological models with perfect fluid in general relativity.

Also the variability of the gravitational constant (G) is a fascinating topic where many physicists and astronomers support the view in Newton's law of gravity and Einstein's general theory of relativity that G is a constant but a few physicists believed that G actually varies with time. A time varying gravitational constant would lead to a modification of the laws of physics with profound implications to astronomy. Dirac (1937) found that in atomic units, G had a value of $10^{-40}$. He interpreted this as G varies inversely with time.

In this letter, we consider the Bianchi type-I bulk viscous polytropic cosmological model using Bianchi type-I metric with variable G and Λ. To obtain the solution of the field equations we assume that $\frac{\dot{A}}{A} = \frac{\dot{B}}{B} = \frac{k_1}{t^n}$ and $\frac{\dot{C}}{C} = \frac{k_2}{t^n}, \eta = \eta_o \rho^\alpha, \bar{p} = k\rho^\gamma$ where $\bar{p}$ is the pressure, ρ is the energy density, η is the coefficient of bulk viscosity, α, k, γ and $\eta_o$ are constants, H is Hubble constant.

2. **Field equations and their solutions:**

We consider the Bianchi type-I space time

$$ds^2 = -dt^2 + A^2 dx^2 + B^2 dy^2 + C^2 dz^2 \qquad (1)$$

where A, B and C are function of t only.

The energy momentum tensor for viscous distribution is given by

$$T_i^j = (\rho + p)v_i v^j - p g_i^j \qquad (2)$$

$$\bar{p} = p - 3\eta H \qquad (3)$$

Where p is the equilibrium pressure, η the coefficient of bulk viscosity and ρ is the energy density with $v_i v^i = -1$.

The Einstein field equations are given by

$$R_i^j - \frac{1}{2} R g_i^j = -8\pi G T_i^j + \Lambda g_i^j \tag{4}$$

where G is the gravitational constant and $\Lambda$ the cosmological constant, which are time dependent.

The expression for scalar of expansion $\theta$ and shear scalar $\sigma$ are

$$\theta = u^i_{;i} = \frac{\dot{A}}{A} + \frac{\dot{B}}{B} + \frac{\dot{C}}{C} = 3H \tag{5}$$

$$\sigma^2 = \frac{1}{2}\sigma_{ij}\sigma^{ij} = \frac{1}{3}\left(\frac{\dot{A}^2}{A^2} + \frac{\dot{B}^2}{B^2} + \frac{\dot{C}^2}{C^2} - \frac{\dot{A}\dot{B}}{AB} - \frac{\dot{B}\dot{C}}{BC} - \frac{\dot{A}\dot{C}}{AC}\right)$$

Einstein's field equation (4) for the metric (1) leads to

$$\frac{\ddot{B}}{B} + \frac{\ddot{C}}{C} + \frac{\dot{B}\dot{C}}{BC} = -8\pi G \bar{p} + \Lambda \tag{6}$$

$$\frac{\ddot{A}}{A} + \frac{\ddot{C}}{C} + \frac{\dot{A}\dot{C}}{AC} = -8\pi G \bar{p} + \Lambda \tag{7}$$

$$\frac{\ddot{A}}{A} + \frac{\ddot{B}}{B} + \frac{\dot{A}\dot{B}}{AB} = -8\pi G \bar{p} + \Lambda \tag{8}$$

$$\frac{\dot{A}\dot{B}}{AB} + \frac{\dot{B}\dot{C}}{BC} + \frac{\dot{A}\dot{C}}{AC} = 8\pi G \rho + \Lambda \tag{9}$$

An additional equation for time changes of G and $\Lambda$ is obtained by the divergence of Einstein tensor i.e.

$$\left(R_i^j - \frac{1}{2} R g_i^j\right)_{;j} = 0 \text{ which leads to}$$

$$8\pi G T_i^j - \Lambda g_i^j{}_{;j} = 0 \text{ yielding}$$

$$8\pi \dot{G} \rho + \dot{\Lambda} + 8\pi G \left\{\dot{\rho} + (\rho + \bar{p})\left(\frac{\dot{A}}{A} + \frac{\dot{B}}{B} + \frac{\dot{C}}{C}\right)\right\} = 0 \tag{10}$$

The conservation of energy Eq. (10) after using Eq. (4) breaks into two equations:

$$\dot{\rho} + (\rho + \bar{p})\left(\frac{\dot{A}}{A} + \frac{\dot{B}}{B} + \frac{\dot{C}}{C}\right) = 0 \tag{11}$$

$$\dot{\Lambda} + 8\pi \dot{G}\rho = 8\pi G\eta \left(\frac{\dot{A}}{A} + \frac{\dot{B}}{B} + \frac{\dot{C}}{C}\right)^2 = 0 \tag{12}$$

### 3. Solution of field equations:

Now, we take

$$\eta = \eta_o \rho^\alpha \tag{13}$$

Where $\eta_o > 0$, $\alpha$=constant.

To obtain the complete solution, we assume the polytropic relation

$$\bar{p} = k\rho^\gamma \tag{14}$$

where k and γ are constants, density ρ is a function of pressure p. [For simplicity we set the constant γ to be unit].

We assume the solution of the system of the equations in the form

$$\frac{\dot{A}}{A} = \frac{\dot{B}}{B} = \frac{k_1}{t^n}, \quad \frac{\dot{C}}{C} = \frac{k_2}{t^n} \tag{15}$$

$k_1$ and $k_2$ are constants [18].

Integrating equation (15), we get

$$A = B = a\exp\left(\frac{k_1 t^{1-n}}{1-n}\right), \quad C = b\exp\left(\frac{k_2 t^{1-n}}{1-n}\right) \tag{16}$$

Where 'a' and 'b' are constants of integration.

Using Eq. (14), (15) in Eq. (11), we get

$$\frac{\dot{\rho}}{\rho} = -\frac{(1+k)(2k_1 + k_2)}{t^n} \tag{17}$$

Integrating, we get

$$\rho = k_3 \exp\left\{-(1+k)(2k_1 + k_2)\frac{t^{1-n}}{1-n}\right\} \tag{18}$$

Differentiating Eq. (18), we get

$$\dot{\rho} = -k_3 \frac{(k+1)(2k_1 + k_2)}{t^n} \exp\left\{-\frac{(k+1)(2k_1 + k_2)}{1-n}t^{1-n}\right\} \tag{19}$$

Using Eq. (9), (15), (16) and differentiating, we get

$$-\frac{2n(k_1^2 + 2k_1 k_2)}{t^{1+2n}} = 8\pi G\dot{\rho} + 8\pi G\eta \left(\frac{2k_1 + k_2}{t^n}\right)^2 \tag{20}$$

Again substituting Eq. (13) and (19) in Eq. (20), we get

$$G = -\frac{n(k_1^2 + 2k_1 k_2)}{4\pi k_3 t^{2n+1}} \exp\left\{\frac{(k+1)(2k_1 + k_2)}{1-n} t^{1-n}\right\}$$

$$\cdot \left[-\frac{(k+1)(2k_1 + k_2)}{t^n} + \eta_o k_3^{\alpha-1} \exp\left\{-\frac{(\alpha-1)(1+k)(2k_1 + k_2)}{1-n} t^{1-n}\right\}\right]^{-1} \tag{21}$$

$$\Lambda = \frac{k_1^2 + 2k_1 k_2}{t^{2n}} + 2n\frac{(k_1^2 + 2k_1 k_2)}{t^{2n+1}}$$

$$\cdot \left[-\frac{(k+1)(2k_1 + k_2)}{t^n} + \eta_o k_3^{\alpha-1} \exp\left\{-\frac{(\alpha-1)(k+1)(2k_1 + k_2)}{1-n} t^{1-n}\right\}\right]^{-1} \tag{22}$$

From Eq. (13) and (18), we get

$$\eta = \eta_o k_3^{\alpha} \exp\left\{-\alpha(1+k)(2k_1 + k_2)\frac{t^{1-n}}{1-n}\right\} \tag{23}$$

where $n \neq 1$.

Thus the metric (1) reduces to

$$ds^2 = -dt^2 + a^2 \exp\left(\frac{2k_1 t^{1-n}}{1-n}\right)(dx^2 + dy^2) + b^2 \exp\left(\frac{2k_2 t^{1-n}}{1-n}\right) dz^2 \tag{24}$$

The density $\rho$, Hubble constant $H_i$, coefficient of bulk viscosity $\eta$, for the model (24)

$$\rho = k_3 \exp\left\{\frac{(1+k)(2k_1 + k_2)}{(n-1)t^{n-1}}\right\} \tag{25}$$

$$H_1 = H_2 = \frac{k_1}{t^n}, H_3 = \frac{k_2}{t^n} \tag{26}$$

where n>1.

$$\eta = \eta_o k_3^{\alpha} \exp\left\{-\alpha(1+k)(2k_1 + k_2)\frac{t^{1-n}}{1-n}\right\}$$

$$\bar{p} = kk_3 \exp\left\{\frac{(1+k)(2k_1 + k_2)}{(n-1)t^{n-1}}\right\}$$

4. **Special cases:**

When $\frac{k_1}{2} = k_2$ and $\alpha = 1$, the Gravitational constant G, Cosmological constant $\Lambda$, Hubble constant $H_i$, the shear scalar $\sigma$ and the expansion scalar $\theta$ for the model (24) are given by

$$G = -\frac{2nk_2^2}{\pi k_3 t^{2n+1}} \exp\left\{\frac{5k_2(k+1)}{1-n} t^{1-n}\right\} \left\{\eta_o - \frac{5k_2(k+1)}{t^n}\right\}^{-1}$$

$$\Lambda = \frac{8k_2^2}{t^{2n}} + \frac{16nk_2^2}{t^{2n+1}} \left\{-\frac{5k_2(k+1)}{t^n} + \eta_o\right\}^{-1}$$

$$H_1 = H_2 = \frac{2k_2}{t^n}, H_3 = \frac{k_2}{t^n}$$

$$\theta = \frac{\dot{A}}{A} + \frac{\dot{B}}{B} + \frac{\dot{C}}{C} = \frac{2k_1 + k_2}{t^n}$$

$$\frac{\dot{G}}{G} \propto H, \Lambda \propto \frac{1}{t^2}, H \propto \frac{1}{t}, G > 0, \rho > 0.$$

The model (24) starts with a big-bang at t=0 when n>0 and the expansion scalar $\theta$ decreases as time t increases. However, when n<0, the expansion in the model increases as the time increases. Also at $t \to 0$, Hubble parameter H and shear scalar $\sigma$ tends to infinity and when $t \to \infty$, Hubble parameter $H_i$ and shear scalar $\sigma$ tends to zero. Since $\lim_{t \to \infty} \frac{\sigma}{\theta} \neq 0$, the model does not approach isotropy for large value of t. As the time t increases, the spacial volume V decreases. The rate of expansion slows down with the increase in time. Since $\eta = \eta_o \rho^\alpha$ and $\alpha > 1$, the model leads to the inflationary phases.

For n=1, Eq. (15) becomes,

$$\frac{\dot{A}}{A} = \frac{\dot{B}}{B} = \frac{k_1}{t}, \frac{\dot{C}}{C} = \frac{k_2}{t}$$

Integrating, we get

$$A = B = k_4 t^{k_1}, C = k_5 t^{k_2}$$

Therefore, metric (1) reduces to

$$ds^2 = -dt^2 + k_4^2 t^{2k_1}(dx^2 + dy^2) + k_5^2 t^{2k_2} dz^2$$

Therefore, the energy density $\rho$, the Gravitational constant G, Cosmological constant $\Lambda$, coefficient of bulk viscosity $\eta$, Hubble constant $H_i$, the shear scalar $\sigma$, the expansion scalar $\theta$ and spacial volume V are given by

$$\rho = N \exp\left\{\frac{(1+k)(2k_1+k_2)^2}{t}\right\}$$

$$\bar{p} = kN \exp\left\{\frac{(1+k)(2k_1+k_2)^2}{t}\right\}$$

where N is the constant of integration.

$$G = \frac{2k_1(k_1+2k_2)t}{8\pi(1+k)(2k_1+k_2)}\left[\exp\left\{\frac{(1+k)(2k_1+k_2)^2}{t}\right\}\right]^{-1}$$

$$\Lambda = k_1(k_1+2k_2)t\left\{\frac{1}{t^3} - \frac{2N}{(1+k)(2k_1+k_2)}\right\}$$

$$\eta = \eta_o N^\alpha \exp\left\{\frac{\alpha(1+k)(2k_1+k_2)^2}{t}\right\}$$

$$H_1 = H_2 = \frac{k_1}{t}, H_3 = \frac{k_2}{t}$$

$$\sigma = \frac{1}{\sqrt{3}}\left(\frac{k_1-k_2}{t}\right) \qquad (27)$$

$$\theta = \frac{\dot{A}}{A} + \frac{\dot{B}}{B} + \frac{\dot{C}}{C} = \frac{2k_1+k_2}{t}$$

$$V = k_4 k_5 t^{(2k_1+k_2)}$$

$$H \propto \frac{1}{t}, \frac{\dot{G}}{G} \propto H, \Lambda \propto \frac{1}{t^2}, G>0, \rho>0$$

As shear scalar σ, expansion scalar θ, Hubble parameter H and cosmological constant Λ are inversely proportional to time. Therefore as the time t increases, shear scalar σ, expansion scalar θ, Hubble parameter H and cosmological constant Λ decreases i.e. the expansion of the universe in the past is large and is becoming slower and slower with the passage of time. This result is in agreement with the observations obtained by many astronomers (Riess et al 1998, 2004, Perlmutter et al 1998 etc.). Also the gravitational constant G and volume of the universe V increases as the time t increases. Since $\frac{\sigma}{\theta} = cons\tan t$ therefore model does not approach isotropy for large value of t. $k_1=k_2$ implies the non-existance of shear in polytropic Bianchi type-I space time.

## 5. Conclusion:

In this paper, we have obtained solutions for the Einstein's general relativity equation in Bianchi type-I space time with polytropic bulk viscous fluid. Here it is observed that when time t→0, then the spacial volume V→∞. this result shows that the universe starts expanding with zero volume and blows up at infinite past and future. The role of bulk viscosity in the cosmic evolution, especially as its early stages seems to be significant. Also when t→0, the energy density ρ, cosmological constant Λ, expansion scalar θ, shear scalar σ, coefficient of bulk viscosity η tends to infinite and when t→∞, ρ, Λ, θ, η, σ tends to zero. Since $\frac{\sigma}{\theta} = cons\tan t$, therefore model does not approach isotropy for large value of t. The spacial volume V increases as time t increases if $k_2 > 0$. Since $\eta = \eta_o \rho^\alpha, \alpha > 0$, the model leads to inflationary solution. The gravitational constant 'G' is increasing as the age of the universe 't' increases over time. An important observation is that gravity was weaker in the past and is becoming stronger as the universe ages which corresponds the results already given by Abdussattar (1997) and others (Vishwakarma 1990, chow 1991, Levitt 1980). The cosmological constant $\Lambda \propto \frac{1}{t^2}$ which follows from the model of Kalligas et al. 1992, Berman et al. 1990, Berman et al. 1989, Bertolami 1986a, 1986b) this form of Λ is physically responsible as observations suggest that Λ is very small in the present universe. Finally the solution presented in the paper are useful or better understanding of the evolution of the universe in Bianchi type-I polytropic bulk viscous cosmological model with variable G and Λ.

**References:**

[1] Allen, S. W. et al.: Mon. Not. R. Astron. Soc. **353,** 457 (2004)

[2] Beesham, A. et al.: Gen. Relat. Gravit. **32,** 471 (2000)

[3] Berman, M. S.: Int. J. Theor. Phys. **29**, 567 (1990)

[4] Beesham, A. et al.: Gen. Relat. Gravit. **21,** 287 (1989)

[5] Bertolami, O.: Nuovo Cimento B **93**, 36 (1986a)

[6] Bertolami, O.: Fortschr. Physics **34**, 829 (1986b)

[7] Chow, T. L.: Nuovo Cimento Lettere **31**, 119 (1981)

[8] Dirac, P. A. M.: Nature **139**, 323 (1937)

[9] Kalligas, D et al.: Gen. Relat. Gravit. **24**, 351 (1992)

[10] Lima, J. A. S., Braz.: J. Phys. **34,** 194 (2004)

[11] Levitt, L. S.: Nuovo Cimento Lettere **29**, 23 (1980)

[12] Padmanabhan, T.: Phys. Rep. **380,** 235 (2003)

[13] Perlmutter, S. et al.: Astrophys. J. **483,** 565 (1997)

[14] Perlmutter, S. et al.: Astrophys. J. **517,** 565 (1999)

[15]  Perlmutter, S. et al.: Nature **391,** 51  (1998)

[16] Peebles, P. J. E. , Ratra, B.: Rev. Mod. Phys. **75,** 559 (2003)

[17] Riess, A. G. et al.: Astron. J. **116,** 1009 (1998)

[18] Reiss, A. G. et al.: Astron. J. **607,** 665 (2004)

[19] Tiwari, R. K., Sharma, Sonia.: Chin. Phys. Lett. **28,** 020401 (2011)

[20] Tiwari, R. K., Sharma, Sonia.: Chin. Phys. Lett. **28,** 090401 (2011)

[21] Tiwari, R. K., Jha, N. K.: Chin. Phys. Lett. **26,** 109804 (2009)

[22] Tiwari, R. K., Dwivedi, U. K.: Fizika B **19,** 1 (2010)

[23] Tiwari, R. K., Rahman, F. and Ray, S.: Int. J. Theor. Phys. **49,** 10 (2010)

[24] Tiwari, R. K.: Astrophys. Astron. **10,** 4 (2010)

[25] Tiwari, R. K.: Astrophys. Space Sci. **318,** 243 (2008)

[26] Tiwari, R. K.: Astrophys. Space Sci. **321,** 447 (2009)

[27] Tiwari, R. K., Tiwari, D. and Shukla, P.: Chin. Phys. Lett. **29,** 010403 (2011)

[28] Tiwari, R. K.: Research in Astron. Astrophys. **10,** 291 (2010)

[29] Vishwakarma, R. G.: Class. Quantum Gravity **17,** 3833 (2000)

[30] Wang, X. X.: Chin. Phys. Lett. **20,** 615 (2003)

[31] Wang, X. X.: Chin. Phys. Lett. **21,** 1205 (2004)

[32] Wang, X. X.: Chin. Phys. Lett. **22,** 29 (2005)

[33] Wang, X. X.: Chin. Phys. Lett. **23,**  1702 (2006)

[34] Zeyauddin M and Saha B.: gr-qc arxiv:1306.2168v1 (2013)

[35] Zhang, H. et. Al.: arxiv:astro-ph/0504178v1 (2005)